\documentclass[10pt,a4paper]{article}

\usepackage{graphics}

\title{Two Combinatorial Models \\
with identical Statics\\
yet different Dynamics}
\author{
David Lancaster\\
\smallskip\\
Harrow School of Computer Science,\\
University of Westminster,\\
Harrow HA1 3TP. UK\\
{\tt lancasd@wmin.ac.uk}
}
\date{2nd July 2003}

\begin{document}

\maketitle

\begin{abstract}
Motivated by the problem of
sorting, we introduce two simple combinatorial models with 
distinct Hamiltonians
yet identical spectra (and hence partition function) and show that 
the local dynamics 
of these models are very different.
After a deep quench, one model slowly relaxes to the sorted state
whereas the other model becomes blocked by the presence of stable
local minima.
\end{abstract}

\section{Introduction}

Viewing optimization problems that arise in computer
science from the perspective of statistical mechanics 
has led to successful insights~\cite{RemiReview}.
From this point of view, where optimization algorithms are
intimately related to the energy or cost function,
the features of the energy landscape are crucial in 
determining the success or otherwise of optimization. 
This connection is
less evident in computer science 
since algorithms do not necessarily have such a direct
connection with the cost function. 
One of the open questions of the field is to understand
the extent to which the performance of algorithms
can be determined by the character of the optimization 
problem and not the details of the algorithm itself~\cite{SearchAlg}.

In this paper we hope to illuminate the issue by
introducing a pair of models with the
unusual feature that although they have the same static properties
determined by the list of energy levels, the energy
landscape is different since the notion of which states 
are close to each other is not the same. 
We perform dynamical simulations of relaxation after a quench
to explore the energy landscape for each of the models
following a similar strategy to that used to investigate
other computer science problems such as satisfiability (K-SAT)
and graph colouring~\cite{SvenNordahl}.

One of the best known optimization models is 
the Traveling Salesman Problem (TSP)~\cite{Percus}.
States can be labeled by permutations
since the paths in the TSP correspond
to various orders of the cities visited. 
The combinatorial models we consider in this paper have states which
are permutations of individual numbers 
rather than permutations of more general
objects such as the set of city coordinates used in the TSP.
In this case the natural optimization problem is
sorting the numbers and we shall take that problem as
sufficient motivation. The problem of sorting 
has been extensively treated in the computer
science literature~\cite{KnuthV3} and many
efficient algorithms are known.
Most of this work is concerned with analyzing the time
taken to perform the sort, though issues such a memory
requirement and ability to use cache are also important.

For some optimization problems, such as TSP, the cost
function is clearly the length of the path 
and this can be taken as the Hamiltonian
of the statistical mechanical model. 
The problem of sorting does not have
an uniquely obvious cost function. 
We require that the
lowest energy corresponds to the sorted state but
have freedom to decide how to measure the degree 
other permutations are sorted.
In this paper we consider two different energy functions that compute
the degree of sortedness in different ways:
one is similar to the TSP and measures the cost in terms
of the length of a path, the other assumes knowledge of the
sorted state and computes the distance from that state in
a direct manner.

By investigating these two models
we discover that the spectrum of energies is identical.
This implies that the partition functions are also
identical and all static properties will be the same.
Yet the Hamiltonians are different, and the 
energy each Hamiltonian assigns to a given state 
is not the same.
We make some preliminary observations on the mapping 
between states with a given energy according to one Hamiltonian
and the same energy according to the other Hamiltonian,
but do not delve into mathematical details.
Of more concern to this presentation is the fact that although
no physical distinction between the two models is visible at
the static level, it is manifest in the dynamics.
We investigate the local dynamics after a deep quench
and show how it displays
very different behavior for each model:
in one case the (sorted) ground state is eventually found, in
the other case it is not.
The difference in behavior is identified as being
due to rather different energy landscapes, with
stable local minima in one Hamiltonian but not the other.

From a physical point of
view, study of the dynamical relaxation after a quench 
falls under the topic of phase ordering kinetics~\cite{Bray}.
From the perspective of computer science, it is the natural
way to study the efficiency of a local search algorithm
in finding the optimum solution to a problem.

\section{Models}

\subsection{State Space}

For statistical mechanical approaches to sorting, the states
or configurations are the permutations of a set of numbers. 
In the following, we shall consider the set of $N$ integers
$1,2\dots N$. A more general model, in which the numbers are taken
to have randomly chosen real values, is briefly described in
appendix \ref{Adisordered}. Most of the properties investigated in this
paper hold for both models, but for the sake of clarity we
shall work exclusively with the simpler integer model.

The states correspond to all possible permutations $P$ of the
set of $N$ integers.
The permutation label can be regarded either passively 
as an ordered $N$-tuple or actively as the mapping
needed to get to that $N$-tuple from the identity~\cite{PCameron}.
Generally we shall employ the first interpretation,
so $P_i$ signifies the $i$'th element 
of the permutation, ($P_1,P_2\dots P_N$),
but the second interpretation will be convenient 
later when we use it to write the permutation in terms of cycles.
We imagine the $P_i$ as analogs of spins on a line, and sometimes we
will refer to this line as a  one dimensional spatial direction.

The size of the state space grows as $N!$
in comparison with typical (Ising)
spin models where the space grows as $2^N$.
It is well-known that models such as this, where the state space
grows faster than exponentially,
have difficulties of interpretation since scaling of
the temperature or energy with $N$ is necessary
to ensure that certain quantities are extensive.
In this work, we do not investigate the phase structure,
and never need to perform this scaling
since we only consider zero (or infinite) temperature.
We refer the reader to M\'ezard and Parisi~\cite{GiorgioA} and to 
Anderson and Fu~\cite{AndersonFu}
for further discussion of this matter.


\subsection{Two Energy Functions}

We consider two energy or cost functions that introduce a measure
of the distance of a permutation away from the identity permutation.
The energy is lowest and vanishes when evaluated for the 
ordered or identity permutation. In writing these expressions
it is convenient to introduce the analog of the Kronecker delta
overlap between individual spins as the
one dimensional distance metric between the numbers $P_i$:
\begin{equation}
d(P_i,P_j) = | P_i - P_j |
\label{eq:1Dmetric}
\end{equation}

The first energy is familiar as the cost function for the 
TSP~\cite{Percus}, here evaluated for the one-dimensional case. 
The boundary conditions are slightly different
from the usual TSP since we insist they be fixed rather than periodic.
The connection with sorting is clearer with fixed boundary conditions
and we need not consider the degeneracy of all states under
cyclic shifts or inversions.
However, it should be noted that the bulk part of the energy
minimizes to either ascending or descending order, and it is
only the boundary conditions that select ascending order.
This energy, which henceforth we shall call the TSP energy ($E_{TSP}$),
is:

\begin{eqnarray}
E_{TSP}(P) & = & {1\over 2 N}\sum_{i=0}^N d(P_{i+1},P_i) - {(N+1)\over 2 N}
\nonumber\\
& = & {1\over 2 N}\left( P_{1} - P_N \right) 
+ {1\over 2 N}\sum_{i=1}^{N-1} | P_{i+1} - P_i | 
\label{eq:Hstring}
\end{eqnarray}

The first form is written in the standard form for TSP and 
we have additionally assumed: $P_0 = 0$, $P_{N+1} = N+1$
for any $P$.
In the second form the constant term is removed, 
by writing the  end terms of the sum explicitly.
Instead of using the function (\ref{eq:1Dmetric}), 
powers (notably quadratic)  
of the node position differences could be considered, but
these energy functions do not appear to be natural in this context and 
do not obey the properties we will demonstrate below.
 
\bigskip

\begin{figure}[ht]
\begin{center}
\includegraphics{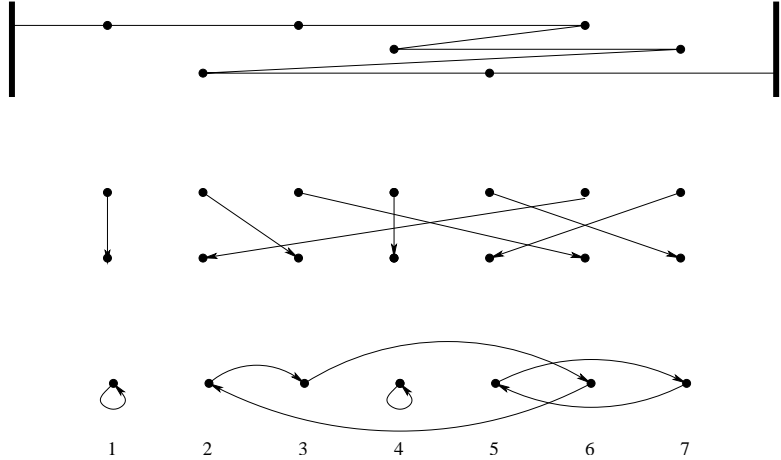}
\caption{
Geometrical representations for an $N=7$ configuration: 
the permutation 1364725.
In the top diagram, the TSP representation is shown.
Note the dependence on boundary conditions
and that there is a continuous path from one wall
to the other.
The second diagram interprets
the displacement energy as bipartite matching: 
each line could be regarded as a spring
pulling the relevant point to its sorted location.
The final diagram is a different representation of the
displacement energy as an assignment problem using just one set of nodes. 
In this representation, the permutation splits
up into several separate pieces corresponding to its
disjoint cycles. 
}
\label{fig:Lines}
\end{center}
\end{figure}

The energy function for our second model, which we shall 
term the displacement energy due to its connection with the
``total displacement'' of R.W.~Floyd\cite{KnuthC}, is
given as a sum of site distances from the sorted configuration
$P^{(0)}_i = i$.

\begin{equation}
E_D(P) = {1\over 2 N}\sum_{i=1}^N d(P^{(0)}_i,P_i)
=  {1\over 2 N}\sum_{i=1}^N | P_i - i |
\label{eq:Hfield}
\end{equation}

This form of cost function is familiar as an ``assignment
problem''~\cite{GiorgioA}.

The choice of distance measure relying on site differences 
is by no means unique. Many other ways of 
defining the distance between configurations are possible.
For example, in the physics literature, an overlap  based on
counting similar links is common~\cite{Overlap}. 
Another overlap (that happens to yield a tractable
model~\cite{Matching}) is
based on the matching problem
and is determined by counting the number of sites that
are in their correct relative positions.

\medskip
The different cost functions can be interpreted geometrically 
as illustrated in figure \ref{fig:Lines}. 
For the permutation, 1364725 with $N=7$, 
used as the example in the figure, the energies may be computed as:
\begin{eqnarray}
E_{TSP} & = & {1\over 2 N}\left(6+2+3+5+6\right) 
- {1\over 2 N}\left(8\right) = 1 
\nonumber\\
E_D & = & {1\over 2 N}\left(1+3+4+2+2\right) = 6/7 
\label{eq:Efigure}
\end{eqnarray}

These expressions and the diagrams in the figure make it clear that 
the energy may be computed as a sum over
runs (either ascending or descending) 
in the case of the TSP energy, and as a sum over cycles 
in the case of the displacement energy. 
A run is the term used in the combinatorial literature for
a subsequence of adjacent elements that are in sorted (or antisorted) order.
For the TSP energy
each contribution is the difference between the maximum
and minimum value contained in the run. 
However, the contribution
of each cycle to the displacement energy is not
usually a single term (for more complex permutations than that
shown in the figure) and it is necessary to look at
runs within a cycle.


\section{Relation between Models}

The energy spectra of $E_{TSP}$ and $E_D$ are identical.
That is, there is a one-one map between the full list of energies
for all states computed with $E_D$ and the list computed with $E_{TSP}$.
Of course there is a shuffling in the way the states are associated
with energies.

\begin{figure}
\begin{center}
\includegraphics{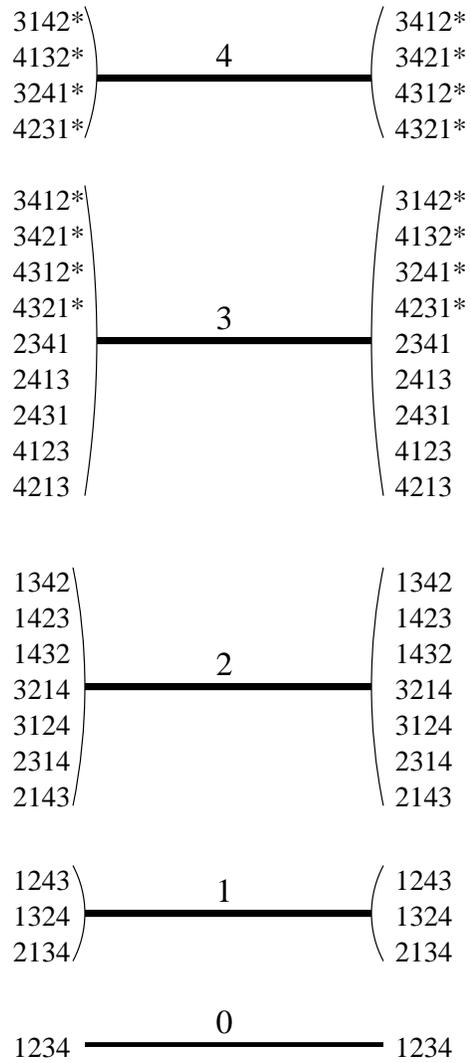}
\caption{
Spectrum for $N=4$. The multiplets and their energies are
shown in the centre 
the left column shows the states
according to the TSP energy and the right column shows them
according to the displacement energy. 
States that are marked with a ``$*$'' 
take different energies depending on the energy function used.}
\label{fig:Spectrum}
\end{center}
\end{figure}
   
For example, figure \ref{fig:Spectrum} shows the spectrum of
the $N=4$ model, with states labeled for each energy function. 
For many, but not all of these $N=4$  
states, the energy is the same
whether evaluated with the TSP or the displacement energy.
The proportion of states with invariant energy decreases at larger $N$.
There are many degeneracies between energy levels:
in the figure there are a 9-plet, a 7-plet, a 4-plet and a triplet
besides the singlet ground state (which is always unique)
that together make the $4! = 24$ states. 
In appendix \ref{Amultordered} 
we list the multiplicity structure 
for small values of $N$. This information might be expected
to lead to an expression for the partition function.
However, we have been unable to obtain a general formula
for these multiplicities and 
according to Knuth\cite{KnuthV3}, the generating function
does not appear to have a simple form.

\subsection{Mapping between Models}

To demonstrate the relation between the two models formally, 
we present a mapping between the states 
$P \to P'$,
(in fact a permutation in the state space) that has the property that
$E_D(P) = E_{TSP}(P')$.

The mapping relies on the representation of the state $P$
in terms of a permutation of the ordered state $(1,2,3\dots N)$
using the cyclic representation.
This mapping is well-known and is for example described in 
Knuth~\cite{KnuthV1}.

We write any permutation in terms of $M$ distinct cycles (singleton
cycles included explicitly). Each cycle is labeled by a superscript
$j$ and contains $m_j$ elements in the cycle as:
\begin{equation}
(i^1_1 i^1_2 i^1_3 \dots i^1_{m_1}) 
(i^2_1 i^2_2 i^2_3 \dots i^2_{m_2}) 
\dots(i^j_1 i^j_2 i^j_3 \dots i^j_{m_j}) 
\dots(i^M_1 i^M_2 i^M_3 \dots i^M_{m_M}) 
\label{eq:cycles}
\end{equation}
We fix the freedom available in the way the cycle form is written
by requiring that: 
\begin{itemize}
\item The largest element appears at the start of 
each cycle: $i^j_1 > i^j_k$ for all other $k$ in the cycle $k = 2,3\dots m_j$. 
\item Cycles appear in the order determined by their first elements:
$i^{j+1}_1 > i^j_1$ for $j = 1,2\dots (M-1)$.
\end{itemize}
Note that this is slightly different from the convention 
defined by Knuth.

The action of this permutation on the ordered state $(1,2,3\dots N)$
is to take the value $i^j_k$ to $i^j_{k+1}$ (or cyclically).
The resulting permutation is the original state $P$ of the map. 
The state $P'$ obtained from the map is given simply by removing
brackets from the definition of the element (\ref{eq:cycles}) above and
regarding the list of numbers as a permutation. 

For example with $N=4$, the cyclic form $(1)(423)$ takes $1234$
to $1342$. Thus $P$ = $1342$ and $P'$ = $1423$. In this case,
both $P$ and $P'$ lie in the same multiplet according to either
the TSP or displacement energy. This is not the case for 
$P$ = $3412$ and $P'$ = $3142$ that appear at the top of figure
\ref{fig:Spectrum} and correspond to the cyclic form $(31)(42)$. 
Nonetheless, $E_D(3412)$ = $E_{TSP}(3142)$.

The displacement energy in the state $P$ 
is given by a sum over cycles as:
\begin{equation}
E_D = {1\over 2N}\sum_{j=1}^M 
\left(  d(i^j_1,i^j_2) + d(i^j_2,i^j_3) 
+ \dots + 
d(i^j_{m_{j-1}},i^j_{m_j}) + d(i^j_{m_j},i^j_1)
\right)
\label{eq:Efield1}
\end{equation}
 
The TSP energy can be evaluated in the state $P'$ to obtain
three terms, one from the elements at the end of the original
sum in (\ref{eq:Hstring}), one from the contribution of each cycle, and
a term corresponding to the intercycle contributions.
\begin{eqnarray}
E_{TSP} & = &
{1\over 2 N}
(i^1_1 - i^M_{m_M})
+
{1\over 2}\sum_{j=1}^{M-1} 
d(i^{j+1}_1,i^j_{m_j})
\nonumber\\
& + &
{1\over 2}\sum_{j=1}^M 
\left( 
d(i^j_1,i^j_2) + d(i^j_2,i^j_3) 
+ \dots +
d(i^j_{m_{j-1}},i^j_{m_j})
\right)
\label{eq:Estring1}
\end{eqnarray}

The cycle sums appearing in  (\ref{eq:Efield1})
and the last term in (\ref{eq:Estring1}) are identical except
for one additional term in the first equation.
The total difference between the energies is given by:
\begin{equation}
E_{TSP} - E_D =
{1\over 2N}
(i^1_1 - i^M_{m_M})
+
{1\over 2}\sum_{j=1}^{M-1} 
| i^{j+1}_1 - i^j_{m_j} |
-
{1\over 2}\sum_{j=1}^M 
| i^j_1 - i^j_{m_j} |
\label{eq:Ediff}
\end{equation}

Now, by using the requirements on ordering of the $i^j_k$'s 
stated above, we find that the 
modulus signs may be removed in each sum and the total vanishes.
This does not occur for other choices of distance function,
but does continue to hold for the model based on real numbers
rather than integers.

The map
is invertible due to the requirements listed above.
Note that the map must leave the ground state unchanged. 
However, the multiplet structure is not preserved, states
that appear in a certain multiplet according to one energy
function may appear in a different multiplet according to
the other energy function. This can be seen in the case of
$N=4$ as the states marked with a ``$*$'' in figure \ref{fig:Spectrum}. 

Several distinct versions of 
the map exist, based on different conventions for the way the
cycle form is written. An approach to understanding the
symmetries of the system would be to
combine a map and the inverse
of a different version. We do not consider this approach
here since it takes us too far afield from
the aim of the present paper.

\subsection{Average Energy}
\label{AvEnergy}

Here we consider statistics of the energy (either TSP or
displacement) spectrum:
the average energy and its fluctuation.
These are averages from a combinatorial
point of view in which all states contribute
equally; thermodynamically they are effectively at 
infinite temperature.

An estimate of the large $N$ behavior is most easily 
obtained by writing the TSP energy in terms of 
$\langle\Delta \rangle$, the average distance between
nodes:
\begin{equation}
\langle E_{TSP}(N) \rangle =
{1\over 2N}\sum_{i=0}^N \langle| P_{i+1} - P_i |\rangle
\approx{1\over 2} \langle\Delta\rangle
={N\over 6}
\label{eq:Eav}
\end{equation}

Where the approximation consists in ignoring correlations 
between the distances between different pairs of nodes.
In this case, the pairs can be imagined as
two independent points thrown at random
in the interval $[0,N]$, so the average distance between
them is $\langle\Delta \rangle = N/3$.

A more formal computation based on the displacement energy
proceeds as follows, where $\sum_P$ indicates a sum over all
permutations.
\begin{eqnarray}
\langle E(N) \rangle & = &
{1\over N!}\sum_{P}{1\over 2N} \sum_{i=1}^N |P_i - i|
\nonumber\\
& = &
{1\over 2 N N!}\sum_{i=1}^N \sum_{P} |P_i - i|
\nonumber\\
& = &
{(N-1)!\over 2 N N!}\sum_{i=1}^N \sum_{k=1}^N |k - i|
\label{eq:ESDavdef}
\end{eqnarray}
The order of summation is exchanged in the second line
after which the sum over permutations becomes simple since there
are $(N-1)!$ permutations in which $P_i$ takes a given value $k$.

The resulting sum can be performed using standard techniques
and a similar argument holds for the second moment.
\begin{eqnarray}
\langle E(N) \rangle & = &
{N^2 -1\over 6N}
\nonumber\\
\langle (E(N) - \langle E\rangle)^2 \rangle & = &
{(N+1)(2N^2 + 7)\over 180 N^2}
\label{eq:ESDav}
\end{eqnarray}

The width of the energy distribution,
$\sqrt{\langle(E-\langle E\rangle)^2\rangle}$,
scales as $N^{1/2}$ and therefore becomes relatively more peaked
at large $N$.


\section{Dynamics}

Although the energy spectra of the two models is identical and
the partition functions are the same, the models still have distinct
properties. This would be apparent by studying the response to
some  external field
that couples to states in the same way in each model.
The displacement energy itself could be regarded as an example
of this kind of additional term in the cost function.
However, in the context of this paper and the problem of sorting,
the distinction is best studied by looking at the
dynamical properties of the models.

We only consider local dynamics: that is the basic
moves are adjacent transpositions.
Of course there is no physical basis
to these models requiring locality, and
in real sorting algorithms non-locality of the 
elementary moves is essential to obtain efficient sorting.
Furthermore, we only consider a dynamics associated with
the energy function - namely the Monte Carlo Metropolis
algorithm. This is natural from a 
theoretical physics perspective
(even though it does not correspond to any physical 
dynamics~\cite{Novotny}),
but algorithms that have much less clear relationship with
the cost function are common in computer science. 
Indeed we could imagine a reasonable
algorithm that selects random sites and transposes
with the neighbor if they are out of order.

The usual approach physicists have used to investigate 
computer science problems is to study the
dynamical relaxation under local search algorithms~\cite{SvenNordahl}.
From the physical point of view, the approach consists
in studying the phase ordering kinetics after a deep quench~\cite{Bray}.

\begin{figure}[ht]
\includegraphics{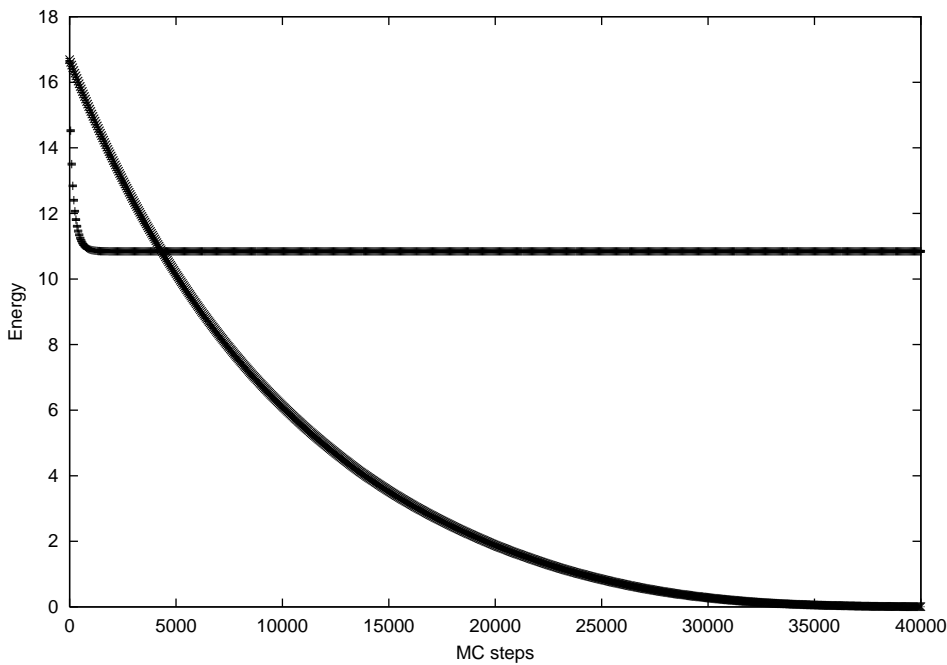}
\caption{
Dynamical evolution of the energy 
according to zero temperature Monte Carlo 
using local moves.
The curve that reaches a plateau is for the TSP energy
and the other is for the displacement energy.  
Size $N=100$. An average over 1000 different
initial states is taken with error bars that are too small to be shown.}
\label{fig:OMonteCarlo}
\end{figure}

\subsection{Zero Temperature Metropolis Algorithm}

The Monte Carlo moves are local adjacent transpositions and 
at zero temperature this effectively 
constitutes a randomized steepest descent algorithm. 
The Metropolis algorithm we shall use 
selects a trial site at random and transposes 
with its (right hand) neighbor provided this move reduces the energy 
(or does not change it).
We perform numerical simulations starting from a random configuration which
typically has energy very close to the average energy computed
in section \ref{AvEnergy}.

Figure \ref{fig:OMonteCarlo} shows the results of 
numerical studies of dynamics according to this algorithm.
The two curves in the figure correspond to dynamics based on 
each of the energy functions we have defined.
Starting from a randomly chosen initial state,
the plot follows the evolution of the energy
averaged over many choices of this initial state.
This situation corresponds to a quench from
very high temperature to zero temperature.

The model based on the displacement energy evolves in the expected
manner: the energy slowly reduces and eventually reaches the
sorted ground state. The other model, based on the TSP
energy starts with a similar initial energy then rapidly reduces to
a plateau value at which level it continues indefinitely. 
No further decrease in energy is evident, even for much longer
runs, and the energy never arrives at the sorted ground state.
This behavior can be improved somewhat by using simulated annealing
but it remains extremely slow and still tends to get stuck.

\begin{figure}[ht]
\includegraphics{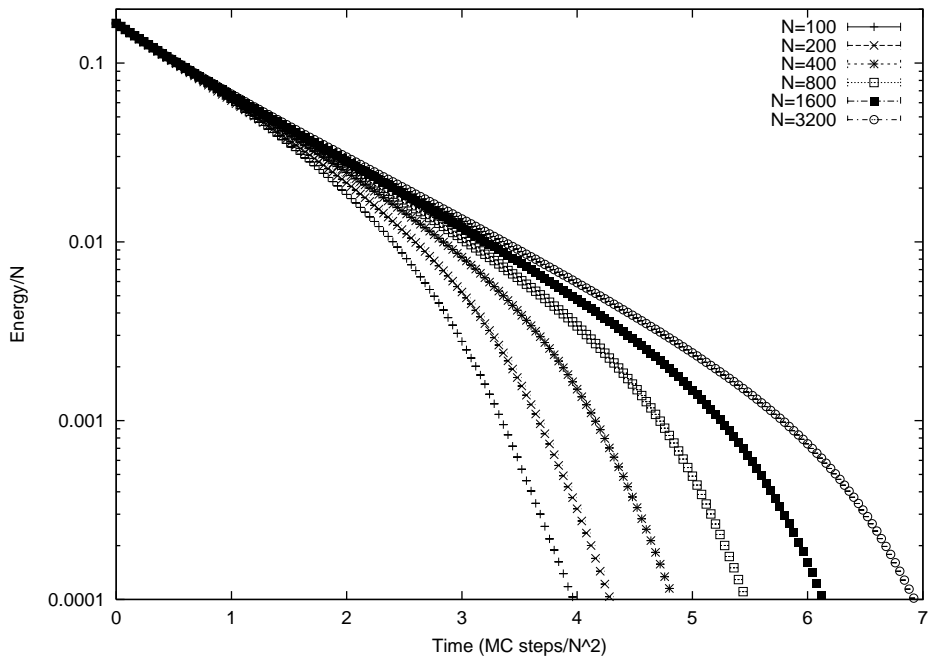}
\caption{
Scaling of the dynamical evolution of the displacement energy 
according to zero temperature Monte Carlo using local moves.
Six size systems from $N=100$ to $N=3200$ are 
shown with axes scaled to demonstrate data collapse.
The energy axis is scaled by $1/N$ and is plotted
logarithmically. The time axis is scaled by $1/N^2$.
An average over 1000 different initial states is taken
with errors that are too small to see on the figure.}
\label{fig:DisplEDynScaling}
\end{figure}

\subsection{Timescales for Energy Decay}
\label{sec:Timescales}

According to figure  \ref{fig:OMonteCarlo}, 
the displacement energy appears to reduce at an
exponential rate. The log-log
plot (not shown) is only able to substantiate this for the first part of the
decay. The quality of this initial exponential decay, and
the form of the subsequent deviations from this form
are shown in  figure
\ref{fig:DisplEDynScaling}. 

Within the initial exponential decay region it is possible to
measure the $N$ dependence of the exponential timescale.
At short times this characteristic time scales as $N^2$ 
(fits using times up to order $N^2$ give the exponent 
with accuracy $\sim 0.1\%$, and value tending 
to 2.0 as fewer points are taken), 
so the time axis
of the figure has been scaled as $t/N^2$ in order to collapse the
data in the initial region.

The deviation from exponential form and absence of data collapse
in the later part of the data is a finite size effect.
With the help of small quantities of data for
very large sizes up to $N=10^5$,
careful measurements of the time required to reach
fixed values of $E/N$ can be made. The way these times scale
with $N$ shows a consistent trend towards an exponent
of 2.0 at larger $N$.

\begin{figure}[ht]
\includegraphics{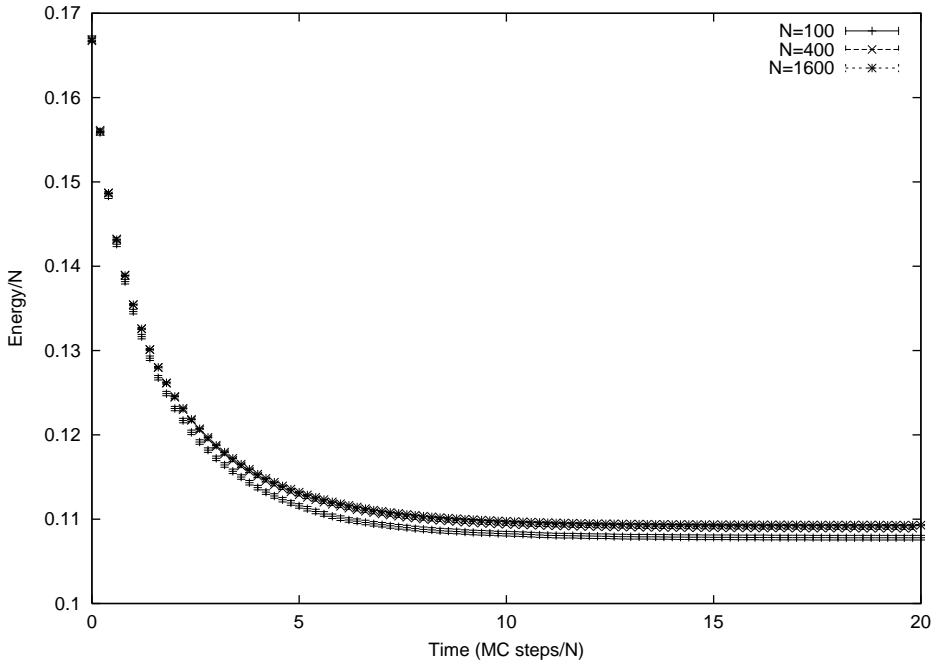}
\caption{
Scaling of the dynamical evolution of the TSP energy 
according to zero temperature Monte Carlo using local moves.
Three size systems, $N=100$, $N=400$, and $N=1600$ are 
shown with axes scaled to demonstrate data collapse.
The energy axis is scaled by $1/N$.
The time axis is scaled by $1/N$.
An average over 1000 different initial states is taken
with errorbars that are the size of the marks.}
\label{fig:StringEDynScaling}
\end{figure}

Certainly for any size that can be simulated, the total time to arrive 
at zero energy is affected by the finite size effects and
has a scaling exponent larger than 2.
Not surprisingly, the resulting sort is rather slower than
achieved with standard sorting algorithms that have best
case behavior increasing as $N\log N$.

For the TSP energy, figure 
\ref{fig:StringEDynScaling} shows a scaled plot for
different size systems.
In this case the time is only scaled by a factor of
$1/N$ and the energy axis is not logarithmic. 
The stability of the plateau energy is very
clear in this figure. A detailed investigation
finds that after the effect of finite size effects
are removed, the per site energy on the plateau is
$E_{TSP}/N = 0.1092\pm0.0001$.

The fact that the evolution timescales of the TSP and
displacement models are respectively $N$ and $\sim N^2$ 
indicates another surprising distinction between the
dynamics of the two models.

\begin{figure}[ht]
\includegraphics{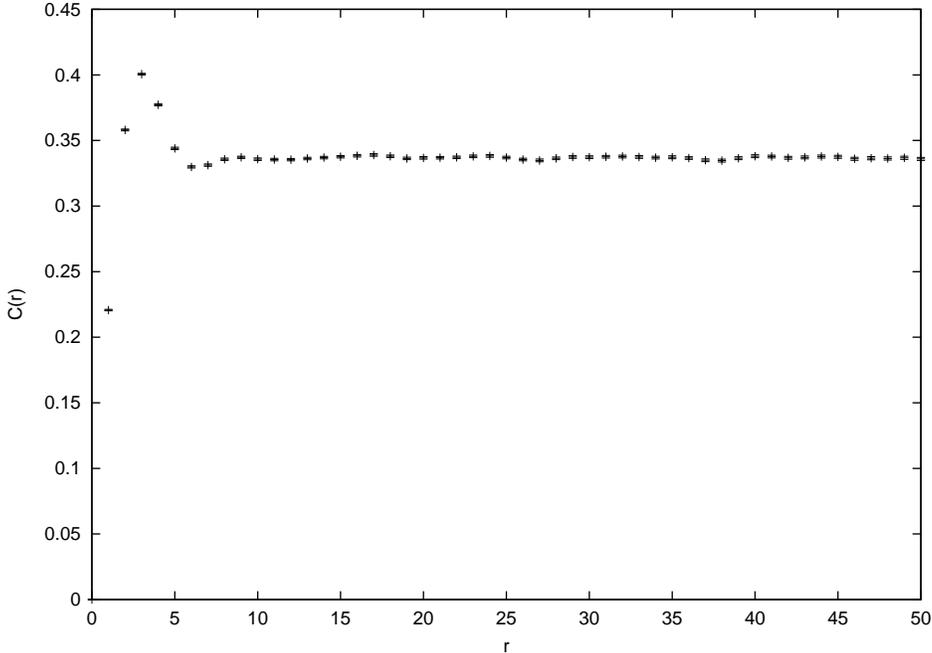}
\caption{
The spatial correlator for configurations chosen
in the plateau region (after 20000 steps on figure 
\ref{fig:OMonteCarlo}, though the form is the same for
all configurations on the plateau)
of the evolution of the TSP energy.
Size  $N=100$ averaged over 
1000 different initial states.}
\label{fig:StringSpatialC}
\end{figure}

\subsection{Spatial Correlations}

Throughout this work we have implicitly considered the
indices to form a line. In this section we consider 
correlations along this line and refer to such correlations
as spatial. Studying the evolution of these correlators
may illuminate the dynamical properties of the model since
dynamical lengthscales may appear.
By analogy with the definition of the distance metric 
(\ref{eq:1Dmetric}), we define the spatial correlator 
at distance $r$ as:

\begin{equation}
C(r) ={1\over N(N-r) }\sum_{i=1}^{N-r} |P_{i+r} - P_i|
\label{eq:Cofr}
\end{equation}

Up to terms relating to boundary conditions,
$C(1)$ is nothing other than the TSP energy.
A small value of $C$
signifies a strong correlation, and in the fully sorted
state it grows linearly $C(r) = r/N$.
From arguments similar to those used to derive the
average energy, it can be shown that the correlator takes
a constant value of $1/3$ when averaged over randomly
chosen configurations (more precisely, 
$\langle C(r) \rangle = (N+1)/3N; r > 0$).
In the figures to follow, we only show $C(r)$ up to
values of $r < N/2$. This is because for larger $r$, only
a small number of pairs appear in the sum (\ref{eq:Cofr})
and the indices of these pairs are often near the
boundaries, thus making this region excessively dependent
on boundary effects.

As the configuration evolves according to the TSP model, the 
spatial correlator remains similar to its initial constant value $1/3$ 
corresponding to the initial random configuration, but some 
structure develops at small $r$. 
Once the plateau is reached, there is no further change in the
spatial correlator and 
figure \ref{fig:StringSpatialC} shows its final form.
The non-trivial structure has a clear size,
of less than 10 units, and neither the size scale nor the 
shape of the structure depend on $N$. 
This scale indicates the distance over which ordering can take
place according to the TSP dynamical process. The mechanism
that blocks further growth of the order beyond this scale is 
discussed below.

\begin{figure}[ht]
\includegraphics{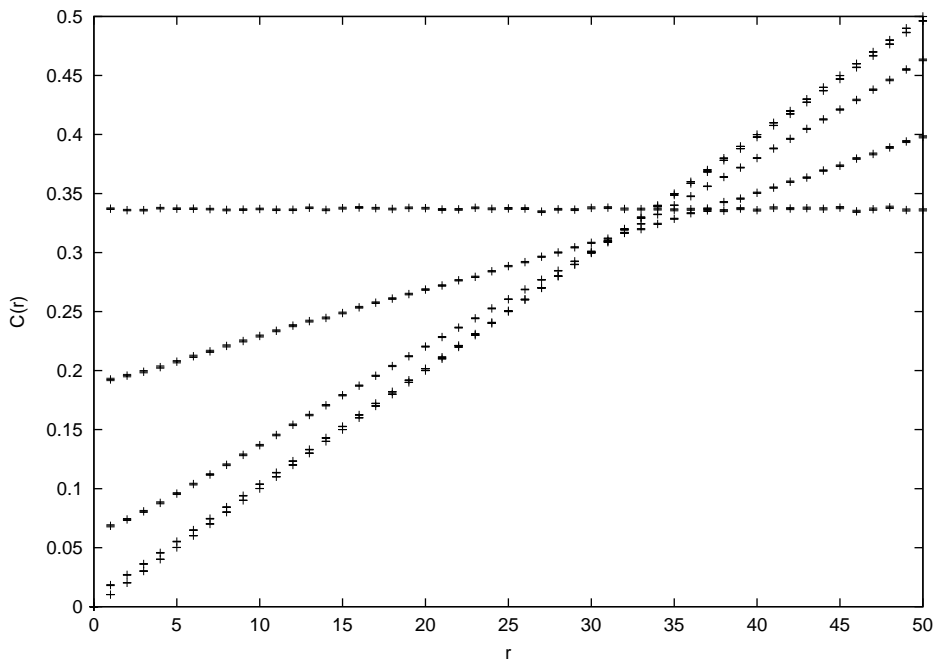}
\caption{
The evolution of the spatial correlator 
for the displacement energy. Size  $N=100$ averaged over 
1000 different initial states.
In order of increasing slope the lines are 
for configurations taken every 10000 Monte Carlo steps 
after the start of the simulation. There is no change
after 40000 steps.
}
\label{fig:DispSpatialC}
\end{figure}

In the case of the displacement energy, the correlator 
evolves as shown in figure \ref{fig:DispSpatialC}. 
Here, no detailed structure
ever appears, and the correlator is always a straight line
with a  gradient that smoothly  evolves towards the
steepest slope corresponding to the final sorted ground state.
There is no characteristic
lengthscale over which ordering takes place and then grows.
It would rather appear that the system becomes organized on 
all scales simultaneously.

\subsection{Time Correlations}

In order to investigate correlations between configurations
at different times, we use the same overlap that was employed for
the displacement energy.

\begin{equation}
C(t,t') ={1\over 2N }\sum_{i=1}^{N} |P(t)_i - P(t')_i|
\label{eq:Ctime}
\end{equation}

Other possibilities are certainly possible. For example,
an overlap based on matching
counts the number of positions where the number has not
changed. The graphs based on this choice do not convey 
any significantly different information from those given below.

\begin{figure}[ht]
\includegraphics{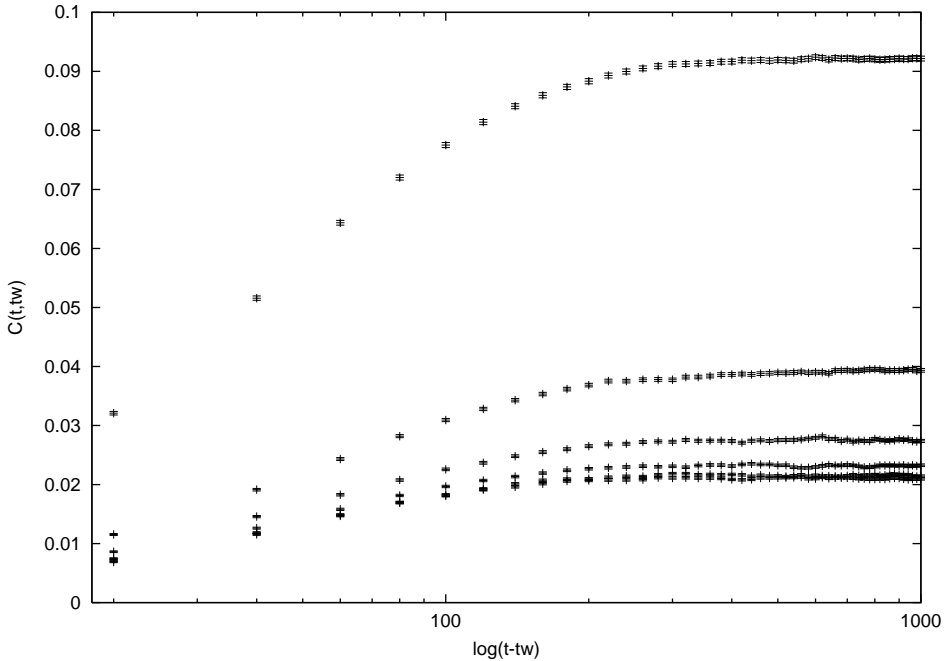}
\caption{
Two-time correlators $C(t,t_w)$ shown for different waiting times
for the evolution according to TSP energy.
Size  $N=100$ averaged over 
1000 different initial states.
Waiting time is zero for the top curve and increases 
by 200 in each lower curve until the bottom curve which
is the same for any waiting time greater than 1000.
This lowest curve indicates that aging does not occur.
}
\label{fig:StringTwoTimeC}
\end{figure}

The simplest correlation to measure is against the
completely sorted ground state. This is none other than
the displacement energy itself, and was shown (for dynamics based on this
energy) in figure \ref{fig:OMonteCarlo}.

Another familiar comparison is against the 
initial state.
For the
dynamics according to the displacement energy, 
the correlation with the 
initial state disappears rapidly (within about $20 N$ steps),
and we do not show a figure in this case.
For the dynamics according to the TSP
energy, we show in figure \ref{fig:StringTwoTimeC}
the two-time correlators for different
waiting time plotted against $\log(t-t_w)$ in the conventional manner.
For $t_w=0$ the correlation against the initial state is included 
in this figure. We have drawn two-time correlators not
because aging occurs in this model - it does not, but 
as a convenient way of demonstrating two features:
that the configurations on the plateau retain some
correlation with the original random state, and to
show that evolution is not frozen on the plateau. 
With this aim, the waiting times shown in figure 
\ref{fig:StringTwoTimeC} are quite short.

The final value of the $t_w = 0$ curve (about $0.092$) is much
less than that associated with correlation between 
random configurations ($1/6$ in this case due to the factor of 2 in the 
definition (\ref{eq:Ctime})), indicating that not
all information in the original configuration has been lost
by the time the plateau is reached. This agrees with the
result of the spatial correlation that indicates that only
local modification takes place. 

The lowest two-time curve holding for all $t_w > 1000$ (for size $N=100$)
corresponds
to waiting times that have reached the plateau.
This curve is not constantly zero as would be the case if there was 
no dynamics happening on the plateau.
The configuration continues to evolve, though the energy 
does not change. 
However, the correlation is bounded: 
irrespective of how long after the waiting time, the
correlator never rises above a certain value (about $0.021$).
This limiting value is independent of $N$.
The reason for this behaviour is the 
limited size of the flat directions that are identified below.

\begin{figure}[ht]
\begin{center}
\includegraphics{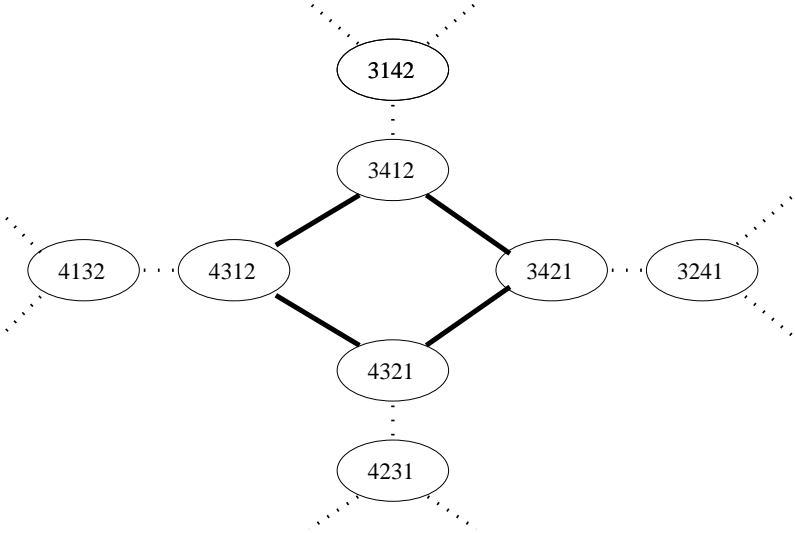}
\caption{
Local minima for $N=4$. The four states in the centre
have TSP energy $E_{TSP} = 3$ while those on the periphery have
$E_{TSP} = 4$. Local transpositions allow three possible
moves between states and these are all shown for the
local minima quartet.}
\label{fig:Localmin}
\end{center}
\end{figure}

\subsection{Local Energy Minima}

The fact that a simple algorithm such as zero temperature
Monte Carlo is unable to find the ground state is hardly
a surprise in optimization models. There are many
statistical mechanical examples of this state of affairs, and
the effect is usually ascribed to the
features of the free energy landscape.
In the K-Satisfiability problem
the difficulty of finding the ground state via a local
search procedure is due to the proliferation of states which
trap the search into metastable phase. Eventually, a large
fluctuation provides a means of reaching the 
ground state~\cite{WalkSAT1,WalkSAT2}.
However, in our TSP model the reason is more prosaic 
and the origin of the effect is due to the presence of
stable local energy minima. 
In this respect, it is similar to the XOR-SAT model, that
also suffers from such minima~\cite{SearchAlg}.

Though the two energy functions have matching energy levels
they have very different energy landscape characteristics. 
The displacement version has no stable local minima, but the
TSP version does. Moreover, these minima appear to proliferate
as $N$ increases and are never avoided.

For small values of $N$ the local minima of the TSP
model may be found explicitly. No such minima exist
for $N=3$. For $N=4$ there are four minima
connected by Monte Carlo moves as shown in figure \ref{fig:Localmin}.
Note that these minima are precisely the states that appear with a ``$*$''
in figure \ref{fig:Spectrum}, and indeed the energy assignments
of figure \ref{fig:Localmin}
are inverted for the displacement energy, so in that case there
are no local minima.

\begin{figure}[ht]
\begin{center}
\includegraphics{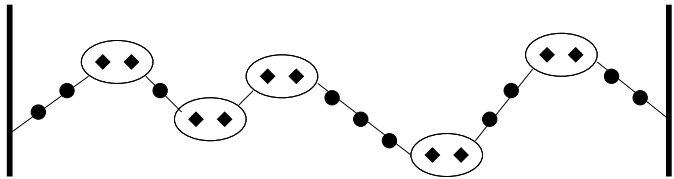}
\caption{
A schematic representation of the trapping configuration
found on the plateau. The turning pairs are shown enclosed.
This configuration could represent the $N=20$ configuration:
5 9 \underline{17 14} 11 \underline{6 8} \underline{20 16} 15 7 4 \underline{1 3} 10 12 \underline{18 19} 13 2.}
\label{fig:PlateauConfig}
\end{center}
\end{figure}

For larger $N$, the trapping configurations appear as
shown in figure \ref{fig:PlateauConfig}. 
They consist of alternating ascending and descending runs
separated by {\it turning pairs}. For an upper turning pair,
each element of the pair is greater than either of the elements
neighbouring the pair, with a similar definition for a lower turning
pair. These  trapping configurations found on the plateau, are
slightly more ordered than random configurations that consist
of ascending and descending runs separated by ordinary turning points.
A turning point  is a requirement on a subsequence of length
3, whereas a tuning pair is a requirement on a longer subsequence of length
4. The dynamical Monte Carlo process performs this small scale
ordering (as observed in the spatial correlators) to arrive at
the plateau configurations. 

The trapping configurations are stable local minima: flat directions
correspond to the moves that interchange the two elements of the
turning pairs, and all other transpositions raise the energy.
The number of states in the trap is therefore $2^{number\ of\ turning\ pairs}$.
Since the number of turning pairs grows linearly with the size of the
system, the size of the trap grows exponentially with $N$.
In appendix \ref{Aplateau}, an estimate for the number of turning
points (same as the number of turning pairs) for plateau configurations
is derived, so the exponential growth is approximately $2^{N/3}$.
Of course, many different traps exist, each with this typical size.

The picture of the trapping configurations in figure \ref{fig:PlateauConfig}
makes the turning pairs appear like domain walls.
This is a reasonable interpretation
since the bulk part of the TSP energy has two different
minima with ascending and descending order and it is these
phases that are separated by the turning pairs. The boundary conditions
raise the degeneracy of the sorted and antisorted phases, but
this effect never comes into play here since the domain walls
are frozen and do not move after the initial relaxation.

The local minima provide a basis for attempting to understand the
value of the plateau energy observed under the dynamics
above. 
A rather naive argument given in appendix
\ref{Aplateau}, based on characterizing the
typical length of ascending or descending runs in the
permutation defining the local minimum state gives the 
per site value of $1/10$. 
This should be compared with the 
numerical value of
$E_{TSP}/N = 0.1092\pm0.0001$ found in section \ref{sec:Timescales}.

\begin{figure}[h]
\begin{center}
\includegraphics{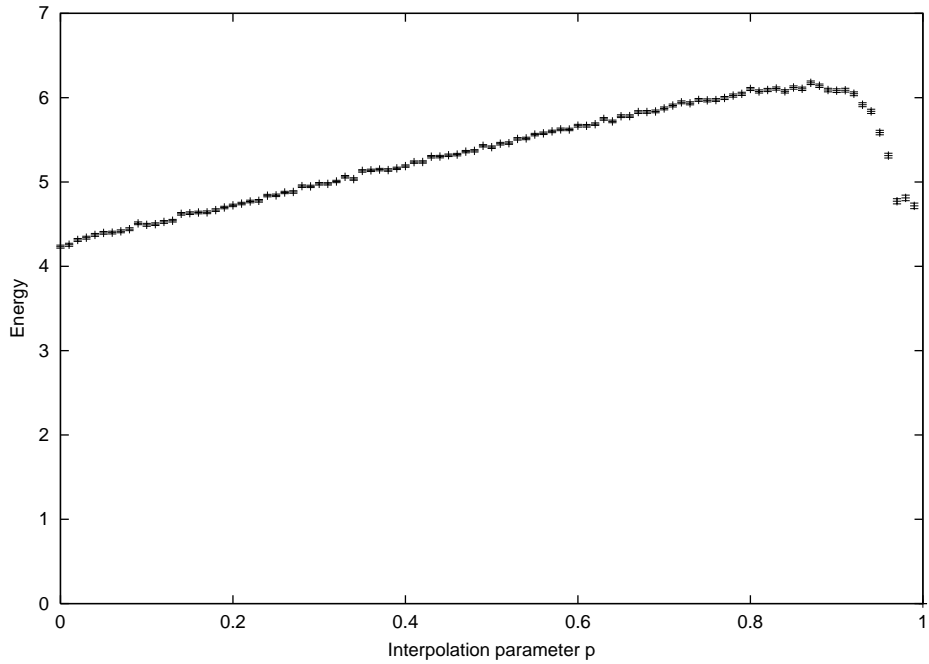}
\caption{
The final (plateau) energy of the interpolating Hamiltonian.
For the ordered model, size $N=40$, with error bars indicating
an average over 1000 random initial states.
Note the ledge near $p=1$, and that only for $p=1$
does the algorithm succeed in finding the zero energy state.}
\label{fig:Interpolate}
\end{center}
\end{figure}

\subsection{Interpolating Model}

Given the rather different dynamic properties
of the two models, it is natural to consider the
behavior of models interpolating between them. 
The interpolating Hamiltonian is:
\begin{equation}
E(p) = (1-p) E_{TSP} + p E_D
\label{eq:Interpolate}
\end{equation}
where $p$ is a parameter in (0,1). 

The average energy (in the combinatoric sense at infinite
temperature) is independent of $p$. All other quantities,
such as the energy of individual states and the degeneracy pattern 
are different from either of the models previously discussed.

Depending on the value of $p$, the dynamics is 
expected to be more 
like one or other of the two lines shown in figure
\ref{fig:OMonteCarlo}.
One might hope for a transition to take place for some intermediate
value of $p$, but in fact the dynamics reaches a plateau for any $p<1$.
Figure \ref{fig:Interpolate} shows how the energy value of the plateau
varies with $p$. For small $N$ 
this curve is composed of a series of steps, ending in a horizontal ledge 
near $p=1$. For larger $N$ the curve becomes smooth and the width of
the ledge shrinks, eventually the curve becomes a sawtooth.

The question arises as to why stable local
minima appear even for an infinitesimal perturbation away from
the displacement energy. Consider the state corresponding to
the completely reversed permutation (with the central two elements
transposed in the case of $N$ even).
The displacement energy of this state is flat
with respect to any of the local moves, though a series of moves
will eventually lead to a lower energy state. 
On the other hand the same state is part of a local minimum
quartet with respect to the TSP energy. 
An infinitesimal addition of the TSP
component is therefore sufficient to make the state
a local minimum, albeit with infinitesimally small barriers.
Although this argument correctly describes the reason that 
trapping can take place for $p$ so close to 1,
the identification of completely reversed states as being
responsible is incorrect since their energy is considerably higher
than the plateau.

\section{Conclusion}

Considering statistical mechanical models based on
states that are permutations of numbers,
we have demonstrated a relationship between two 
models with distinct energy functions.
Statically their partition functions are identical since
the energy spectrum is the same. Dynamically there are
substantial differences since one model has local energy minima
and the other does not.
This strange situation, that models with identical static
properties have distinct local dynamics, is not a paradox. 
In this case it is clearly due to the shuffling of states
that modifies the energy landscape by 
rearranging the energies of states that are close to each other.

The static analysis showed that the energy spectrum
was the same by demonstrating a one-to-one map relating 
states with the same energy in each model. This map
continues to hold for the more general model based
on real numbers rather than integers.

Dynamically, we simulated a deep
quench and showed that
the energy decay proceeds in a completely different
way in each of the two models. For the displacement model,
no dynamical lengthscale appears as the system is  
reorganised. Characteristic timescales do appear
and vary as $N^2$, though with strong finite size effects.
For the TSP model, there was rapid (timescale varying as
$N$) decay to a plateau as a result of reorganisation over
small spatial scales of size less than 10 units, that did not completely
destroy the corelation with the initial state.
Evolution continued to occur on the plateau, but was
limited in range. This was interpreted in terms of
trapping configurations with turning pairs.
Flat movement within the trap was still possible
between about $2^{N/3}$ local minima states.

Optimization problems of most interest, such as K-Satisfiability,
are much richer than the models presented here. Nevertheless,
in the context of the one parameter family of interpolating
models, we have shown that our simple local search procedure 
exhibits a transition
(at the very edge of the interpolation region) from being able to
sort the numbers to becoming trapped by local minima. 
We hope that these models  provide a simple
arena for studying the general question of the role of the
energy landscape in the performance of local search 
algorithms.

\newpage
\appendix

\section{Appendix: Model based on Real Numbers}
\label{Adisordered}

Instead of the integer model discussed in the text,
this more general model is based on 
the set of $N$ real numbers $x_i$, $i=1,2\dots N$, with each  
$x_i$ in the range $[0,N+1]$. 
Without any loss of generality, we take the $x_i$'s to be
ordered ($x_i < x_j$ for all $i < j$) 
as this makes the identity permutation of the indices correspond to
the lowest energy, or sorted state. 
The $x_i$'s should be regarded as quenched random variables, 
and for this reason the model might be regarded as a disordered
model in contrast to the integer ordered model. 
The disorder however, is not of the independent
variety that has been considered for both assignment and TSP problems
using a replica approach
in \cite{GiorgioA,GiorgioB}, but rather corresponds to Euclidean
distances in one dimension. Here, the replica approach becomes
intractable since the average over disorder couples sites.

For the disordered model, the definitions of the two energy
functions remain exactly as given on the first lines of equations
(\ref{eq:Hstring}) and (\ref{eq:Hfield}), however the distance
function is replaced by:
\begin{equation}
d(P_i,P_j) = | x(P_i) - x(P_j) |
\label{eq:1Dmetricdisordered}
\end{equation}
Irrespective of the choice of $x$'s, there are degeneracies 
in the spectrum, but not to the extent found in the ordered model.
For example, for $N=4$, the disordered model has energy levels
with multiplicities (9,4,3,3,1,1,1,1,1). In the ordered model
these combine to give (9,7,4,3,1).

The mapping between the energy levels according to each model
continues to hold, however there is a small difference in 
the moments. Here the moments are defined after
additional averaging over the $x$'s. 
The mean energy is the same as in the ordered model by design - indeed
it was this criterion that selected the range of the
$x_i$'s to be $[0,N+1]$. The second moment is however slightly larger
than quoted in (\ref{eq:ESDav}).

\begin{equation}
\overline{\langle (E(N) - \langle E\rangle)^2 \rangle}  = 
{(N+1)^2(3N^2 + 10N + 2)\over (N+2) 180 N^2}
\label{eq:ESDdisav}
\end{equation}

Even for individual instances of the disordered model, 
similar dynamical behaviour to that described in the text
for the integer model is observed. After averaging
over realizations of the $x_i$'s, the similarity becomes even closer.

\section{Appendix: Multiplicities of Degeneracies}
\label{Amultordered}

The multiplicities are most easily obtained 
using the displacement energy $E_D$
This may be evaluated for every permutation 
and $D_N(E)$ denotes the number of times that 
displacement $E$ appears. For small $N$ (up to $N \approx 16$),  it is 
straightforward to numerically enumerate these coefficients,
and the first few are given in the table below.

\begin{center}
\begin{tabular}{|c|c|c|c|c|c|c|c|c|c|c|}
\hline
$N$ & $D(0)$ & $D(1)$ & $D(2)$& $D(3)$& $D(4)$& $D(5)$
& $D(6)$ & $D(7)$ & $D(8)$& $D(9)$ \\
\hline
1 & 1 & 0 & 0 & 0 & 0 & 0 & 0 & 0 & 0 & 0  \\
2 & 1 & 1 & 0 & 0 & 0 & 0 & 0 & 0 & 0 & 0  \\
3 & 1 & 2 & 3 & 0 & 0 & 0 & 0 & 0 & 0 & 0  \\
4 & 1 & 3 & 7 & 9 & 4 & 0 & 0 & 0 & 0 & 0  \\
5 & 1 & 4 & 12& 24& 35& 24& 20& 0 & 0 & 0  \\
6 & 1 & 5 & 18& 46& 93&137&148&136&100& 36 \\
\hline
\end{tabular}
\end{center}
\bigskip  

The usual arguments intended to lead to recurrence relations 
that would relate  $D_{N+1}(E)$ to sums of $D_N({E'})$ are not helpful
in this case. Although there are some relations based on 
a sum over partitions of $E_S$, these are only valid in the
region below the diagonal of the table. 
The problem appears to be non-trivial
and indeed, according to Knuth\cite{KnuthV3}, the generating function
does not appear to have a simple form.






\section{Appendix: TSP Plateau Energy}
\label{Aplateau}

The average TSP energy is estimated after the effect
of the Monte Carlo moves. It is convenient to evaluate
the TSP energy by summing contributions from
ascending and descending runs, rather than from
adding each individual term which leads to many cancellations. 
The theory of
runs of this kind is presented in \cite{BartonDavid},
but very little of the general development is necessary here.
The main property we use is that their endpoints are characterized
by turning points in the permutation sequence.
We assume that $N$ is large and only consider
leading terms.

\begin{equation}
\langle E_{TSP} \rangle \simeq
{1\over 2N}\sum_{i=0}^N \langle| P_{i+1} - P_i |\rangle
\approx {1\over 2N}\langle s\rangle \langle \Delta\rangle
\label{eq:Erun}
\end{equation}

Where $\langle s\rangle$ is the average number of runs,
and $\langle \Delta\rangle$ is the average (absolute) change 
between the start and end of the run
($\Delta = d(P_{start},P_{end})$). 
The first approximation in this approach 
is to ignore the correlations between the number of runs and
the value of $\Delta$ for that run in the formula above. 

To illustrate this form, let us use it to reproduce the
average energy of a random configuration. In that case,
$\langle s\rangle = 2N/3$, since by considering sets of
three adjacent points, the central one has a maximum or minimum value in
4 out of the 6 equally likely orderings. To deduce $\langle \Delta\rangle$
we consider the average value taken at upper turning points
separating runs. Since
$\langle max(P_1,P_2,P_3)\rangle = 3N/4$ and a symmetric
result for the minimum, 
$\langle \Delta\rangle = N/2$. Combining these results
in equation (\ref{eq:Erun}) reproduces
$\langle E_{TSP}(N) \rangle = N/6$ as was derived in the text
by considering contributions from all neighboring pairs.

\begin{table}[h]
\bigskip
\begin{center}
\begin{tabular}{|c|c|c||c|c|c||c|c|c|}
\hline
P & $\Delta E$ & $\Delta s$  &P & $\Delta E$ & $\Delta s$  & P & $\Delta E$ & $\Delta s$  \\
\hline
1234 & + & 2  & 1432 & 0 & 0   & 4312 & + & 1 \\
2134 & + & 1  & 1423 & - & -1  & 3421 & + & 1 \\
1324 & - & -2 & 1342 & 0 & 0   & 3412 & + & 0 \\
1243 & + & 1  & 4231 & - & -2  & 4213 & 0 & 0 \\
2143 & + & 0  & 3241 & - & -1  & 4123 & 0 & 0 \\
3214 & 0 & 0  & 4132 & - & -1  & 2431 & 0 & 0 \\
3124 & 0 & 0  & 3142 & - & 0   & 2341 & 0 & 0 \\
2314 & - & -1 & 4321 & + & 2   & 2413 & - & 0 \\
\hline
\end{tabular}
\caption{
Effect of transposing central pair of 4 points.  $\Delta E$ indicates
the sign of the energy change as a result of this move.  $\Delta s$
shows the change in number of (internal) turning points. }
\label{tab:MCN4}
\end{center}
\end{table}

An estimate of the value of the average in the plateau is obtained by
taking into account the effect of the Monte Carlo moves. 
We consider a subsequence of four points
and look at the effect of a transposition on the central
pair of the four, in all 24 possible orderings.
The four points are labeled 1234, but this only signifies 
their relative magnitude.
Table \ref{tab:MCN4} indicates whether the transposition
causes a positive, negative or zero change in energy 
and lists the change in
number of (internal) maxima or minima.

Using this table, we find that moves that do not
change the energy do not alter the number of runs.
Moreover, there are six configuration that
both reduce the energy and number of runs. 
These are the pair 1324, 4231, and the quartet
2314, 1423, 3241, 4132.
If we take their naive weights from the initial
random configuration, then such transpositions for every
pair of adjacent points leads to an 
average reduction in
$\langle s\rangle$ of $2\times N\times 2/24 + 1\times N\times 4/24$.
So the final value after one Monte Carlo sweep is 
$\langle s\rangle = N/3$.

The argument for the average value at a maximum
is extended from 3 (given in the example above) 
to 4 points, where it reproduces the
3 point case when all contributions are included.
However, some of these contributions are removed
by the Monte Carlo move (we only include those configurations
with $\Delta E \ge 0$ in table \ref{tab:MCN4}) resulting in
$\langle \Delta\rangle = 3N/5$. 

Overall we obtain the estimate 
$\langle E_{TSP}(N) \rangle = 1/2N \times N/3 \times 3N/5 = N/10$,
to be compared with the numerical value of 
$E/N = 0.1092\pm0.0001$.
The approximations in this approach are quite drastic.
First we ignored correlations between the number of runs and
the value of $\Delta$ for that run in formula (\ref{eq:Erun}).
Then we just looked at the effect of one Monte Carlo sweep,
used naive weights and ignored any influence of
one Monte Carlo move on another.
It is therefore surprising that we obtain such
a reasonable estimate.
Indeed, numerical studies show that the estimates of both 
$\langle s\rangle$ and $\langle \Delta\rangle$ are
incorrect by about 10\%.

\newpage


\begin{thebibliography}{99}
\bibitem{Matching}
The generating function for the matching problem
is given in \cite{KnuthV3}.
\end{thebibliography}
\end{document}